\RequirePackage{fix-cm}
\documentclass[twocolumn,epjc3]{svjour3}  
\usepackage[numbers,sort&compress]{natbib}
\usepackage{subcaption}

\smartqed  % flush right qed marks, e.g. at end of proof
\RequirePackage{graphicx}
\usepackage{xcolor}
\usepackage{booktabs}
\usepackage{xurl}
\usepackage[hidelinks]{hyperref}
\usepackage{orcidlink}
\usepackage{float}
\usepackage{dblfloatfix}
\usepackage{amsmath} % for align and equation environments
\newcommand{\opticks}{\texttt{Opticks}}
\newcommand{\gfour}{\texttt{GEANT4}}
\newcommand{\larsoft}{\texttt{LArSoft}}
\journalname{Eur. Phys. J. C}
\begin{document}
%\linenumbers

\title{Accelerating Optical Photon Simulation in DUNE with Opticks}


\author{I. Parmaksiz\thanksref{e1,addr1}\orcidlink{0009-0002-1478-3977} 
        \and
        A. Higuera\thanksref{addr1}\orcidlink{0000-0001-9310-2994} %etc.
        \and 
        L.~Paulucci\thanksref{addr2}\orcidlink{0000-0002-1041-6064}
        \and
        V.~Pec\thanksref{addr3}\orcidlink{0000-0003-4104-829X}
        \and
        E. Forino\thanksref{addr1}
}

\thankstext{e1}{e-mail: ip33@rice.edu}
\institute{Rice University, Houston, TX 77005, USA \label{addr1} \and 
 Instituto Tecnológico de Aeronáutica, Brazil, Sao Jose dos Campos, Brazil \label{addr2}   \and 
 Institute of Physics, Czech Academy of Sciences, 18200 Prague 8, Czech Republic \label{addr3}
}

\date{}

\maketitle
\abstract{Optical photon simulation is among the most computationally demanding tasks in complex and large detector geometries. In the Deep Underground Neutrino Experiment (DUNE), the scale of the far-detector modules makes photon-by-photon transport with \gfour~prohibitively expensive on CPUs. We present the first implementation and performance evaluation of GPU-accelerated optical photon simulation at the 10~kt scale, based on \opticks~and applied to the DUNE far-detector horizontal-drift (FD-HD) geometry. On the full FD-HD geometry, \opticks~propagates the same photons as \gfour~with a speedup of $313 \pm 3$ over single-threaded and $83 \pm 1$ over four-thread \gfour, and is validated against the reference \gfour~simulation across all metrics considered. We further integrate \opticks~into the \texttt{LArSoft}-based DUNE software stack. Photons simulated on the GPU with \gfour-equivalent physics retain full Monte Carlo fidelity at a computational cost that makes high-statistics optical studies and the generation of labeled datasets for machine learning feasible at the kiloton scale.
}
%\keywords{Optical simulations \and DUNE \and \opticks}

\section{Introduction}
 
In the modern landscape of high-energy physics (HEP), the creation of a digital twin, commonly known as detector simulation, is among the most computationally demanding
tasks an experiment undertakes, particularly simulations within the
\gfour~framework~\cite{AGOSTINELLI2003250,1610988,ALLISON2016186}. In some instance, detector
simulation can account for as much as 40\% to 80\% of an experiment's total CPU usage~\cite{thedunecollaboration2022dune,Li:2023ocy,ATLAS:2020pnm,Althueser_2022,HEPSoftwareFoundation:2017ggl}. These cover signal and background processes across an enormous dynamic range of energies from TeV-scale muons down to eV-scale optical photons traversing meters of complex materials. This computational burden is projected to grow substantially and has been identified as one of the principal computing bottlenecks of the coming decade in the HEP community~\cite{HEPSoftwareFoundation:2017ggl,Calafiura:2729668}. 

Optical photon simulation is typically the most computationally expensive part of the workflow, as optical photons far outnumber other secondaries. Each must be individually propagated through the detector, undergoing scattering, reflection, and wavelength shifting before ultimately being absorbed or detected.
The cost of this photon-by-photon transport scales with both the photon yield of the medium and the size of the volume that must be traversed. DUNE, a next-generation long-baseline neutrino experiment currently under construction, sits at the extreme of both. Each of its far-detector modules is housed in a cryostat of approximately $14 \times 12 \times 58$~m$^{3}$~\cite{DUNE:2020txw}, and a single neutrino interaction can produce on the order of $10^{7}$ optical photons~\cite{DUNE:2020lwj,DUNE:2020ypp}.

%DUNE's physics program centers on precision measurements of neutrino oscillations and to test the three-flavor oscillation paradigm with unprecedented precision, in addition to a program that includes the detection of neutrinos from core-collapse supernova~\cite{DUNE:2020zfm}, the study of atmospheric neutrinos~\cite{DUNE:2026yly}, and an abroad portfolio of beyond-the-Standard-Model (BSM) searches~\cite{DUNE:2020fgq}. DUNE's first two far detectors will be composed of liquid argon time projection chambers (LArTPCs), each 10~kt fiducial. Liquid argon is a prolific scintillator, emitting vacuum ultraviolet (VUV) light at 128~nm at a yield of roughly $2.4\times10^{4}$ photons

%per MeV of deposited energy at the nominal drift
%field~\cite{DUNE:2020txw,Doke:1990rza}.

%; a single neutrino interaction $\mathcal{O}(1~\mathrm{GeV})$ is therefore expected to produce on the order of $10^{7}$ optical photons. %To detect these optical photons, a set of photon detectors (PDs) is a key component of DUNE far detector LArTPCs, the scintillation light collected by the PDs provides the absolute event time ($t_{0}$), enabling triggering and drift-coordinate localization of non-beam events and contributing complementary calorimetric information to the TPC charge signal. 

Optical transport of the resulting billions of photons with single-threaded \gfour~ on CPUs is, however, prohibitively expensive at these volumes. To overcome this, experiments have adopted approximate methods. One widely used example is the lookup-table, or ``photon library'' approach, which precomputes detection probabilities from voxelized photon scans of the detector volume~\cite{Marinho_2022}. More recent semi-analytical models instead predict the number and arrival-time distribution of detected photons directly from the relative distance of the energy deposit and the photosensor~\cite{Garcia-Gamez:2020mmf}. Machine learning (ML) approaches include generative and neural-network photon-transport surrogates~\cite{Mu_2022,lei2022}.  While useful for a large-volume detector, these fast approaches share a fundamental limitation: they, by construction, bypass the underlying particle transport and therefore cannot provide complete truth-level information. In a full simulation, each photon is propagated step-by-step through the detector geometry, naturally incorporating stochastic fluctuations in Rayleigh scattering, bulk absorption, and boundary reflections, leading to path-length delays. In contrast, fast approaches rely on spatially averaged detection probabilities and reconstruct arrival times using parameterized probability density functions. As a result, the fast approach produces an averaged approximation; it fails to reproduce localized optical propagation effects, smears out correlated timing structures across neighboring channels, and limits knowledge of the truth-level parentage connecting an individual detected photon to its exact trajectory and micro-interaction in the material, which is essential for developing and validating low-level reconstruction, optical triggering, and calibration algorithms.
 
The gap between fidelity and computational cost has motivated community-wide efforts to accelerate optical-photon transport simulations on modern parallel hardware. \gfour~itself introduced event-level multithreading~\cite{ALLISON2016186}. More recently, attention has shifted to GPUs, the Celeritas project offloads full electromagnetic shower simulation to GPUs within \gfour-driven workflows, achieving order-of-magnitude speedups ~\cite{Johnson:2024z022,refId0} and has plans to incorporate optical photons. For the specific task of optical photo transport, an open-source API called \opticks~\cite{Blyth:2017pjs,Blyth:2024lbv} integrates \gfour~ with the NVIDIA OptiX ray-tracing engine~\cite{Parker10OptiX}, automatically translating the \gfour~ geometry to the GPU and implementing the optical physics processes, including Rayleigh scattering, absorption, reemission, and boundary interactions in CUDA~\cite{cuda}. \opticks~has demonstrated speedups exceeding a factor of 1500 relative to single-threaded \gfour~ for the JUNO liquid-scintillator detector~\cite{Blyth:2024lbv}, and has been explored for application in the LHCb RICH detectors~\cite{Li:2023ocy}, adopted for optical time projection chamber (TPC) R\&D~\cite{NEXT:2025opticks} and used on idealized large liquid argon (LAr) TPC~\cite{galgoczi2026gpuopticalphotonmonte}.

\opticks~ currently supports the majority of \gfour~ solids. However, certain solids, such as \texttt{G4TessellatedSolid}, and the wavelength‑shifting (WLS) physics process are not yet implemented. Work is planned to extend \opticks~ with these missing capabilities. 
In parallel, \texttt{Simphony}~\cite{galgoczi2026gpuopticalphotonmonte}, built on \opticks, has introduced a GPU‑accelerated implementation of the \gfour~ \texttt{G4OpWLS} model, which has been validated in a simplified liquid argon TPC geometry. Preliminary results suggest that the inclusion of WLS physics can be achieved without compromising \opticks’ performance~\cite{galgoczi2026gpuopticalphotonmonte}.

In this paper, we present the first implementation and performance evaluation of \opticks~for optical photon simulation in the full 10~kt horizontal-drift far-detector geometry of DUNE (see Fig.~\ref{fig:fd_hd}), demonstrating GPU-accelerated, full-fidelity photon transport at the scale of a 10 kt detector module.

\section{DUNE HD far detector geometry }
\label{ref:sec2}
\begin{figure}[ht!]
  \includegraphics[width=0.48\textwidth]{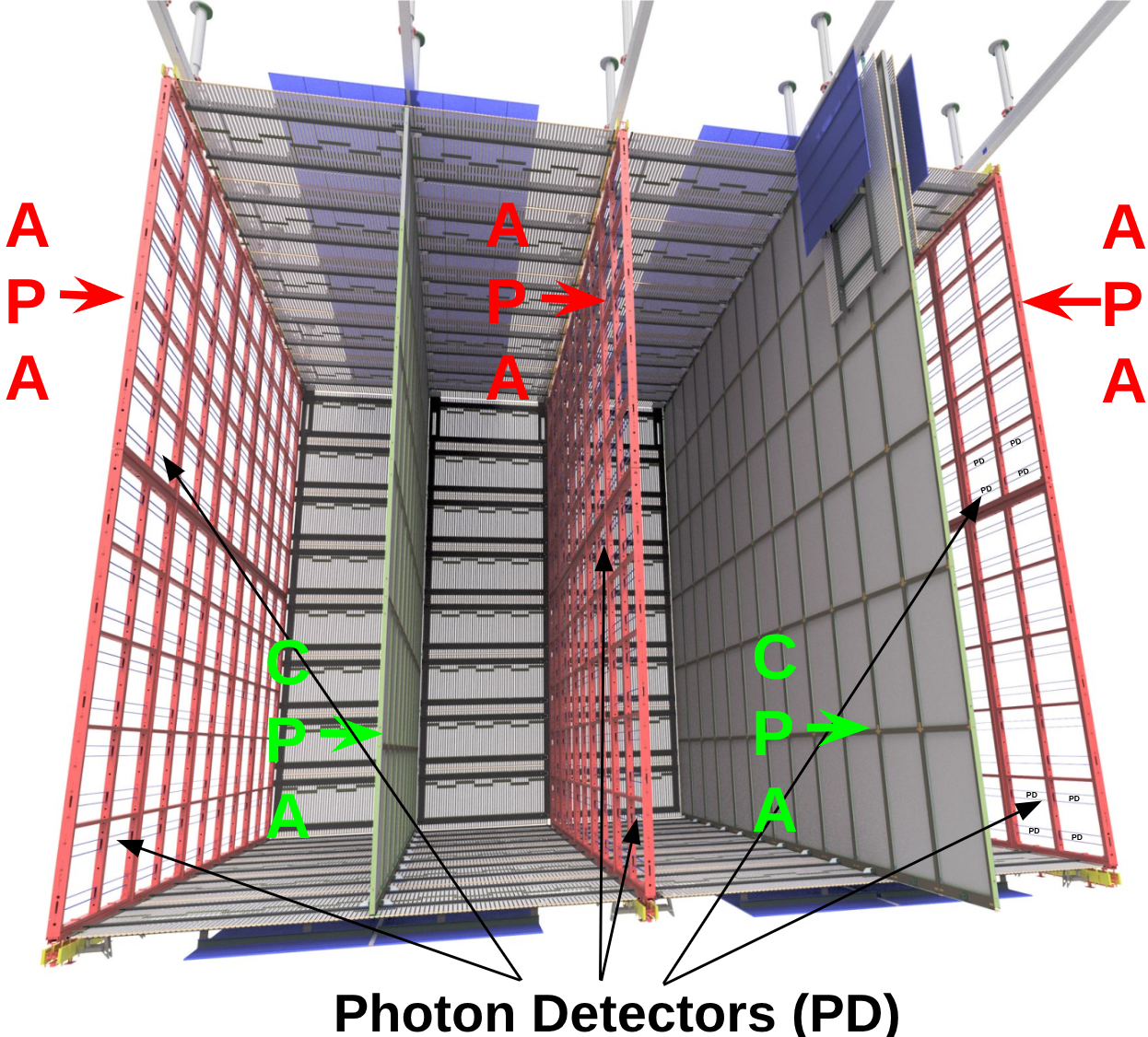}
\caption{Layout of the 10~kt FD‑HD module illustrating the Anode Plane Assembly (APA) arrays, Photon Detectors (PDs), and Cathode Plane Assembly (CPA) arrays. Adapted from~\cite{DUNE:2020txw}.  }
\label{fig:fd_hd}  
\end{figure}

DUNE's primary physics program centers on precision measurements of the rates and spectra of $\nu_\mu \rightarrow \nu_e$ and $\bar{\nu}_\mu \rightarrow \bar{\nu}_e$ oscillations. Identifying these signatures among copious backgrounds requires a detector that combines a fiducial mass of many kilotons with sub-centimeter spatial resolution, a combination uniquely provided by the LArTPC technology. The first two DUNE far-detector modules are LArTPCs of 10~kt fiducial mass each. LArTPC collects the ionization electrons liberated by charged particles traversing the argon on a readout plane. In one of the DUNE modules, the horizontal drift module (FD-HD), the electrons drift horizontally under the uniform electric field established between the vertically oriented cathode and anode planes, with the active volume enclosed by a field cage.
Each anode wall consists of anode plane assemblies (APAs) with wire planes wrapped on insulating frames. As ionization electrons drift to the APA, the two outer planes record induced signals, while the innermost collects charge. PDs are integrated into the APAs: ten per APA, spaced 592~mm apart, totaling 1500. Of these, 500 in central frames collect light from both directions, while 1000 near cryostat walls collect from one direction~\cite{DUNE:2020txw}, see Fig~\ref{fig:fd_hd} for a layout representation of the geometry.
In the FD-HD design, the PD system is implemented with the X-ARAPUCA technology~\cite{Machado:2018rfb}, in which incident VUV scintillation photons are wavelength-shifted and trapped by a combination of dichroic filters and a wavelength-shifting plate inside a highly reflective enclosure until they are detected by silicon photomultipliers (SiPMs) mounted along the module edges. The initial calibration of X-ARAPUCAs for the FD-HD has been experimentally conducted and documented in prior reports~\cite{Palomares_2023,Alvarez_XArapuca}. Dedicated \gfour~simulations of scintillation photon detection with a single X-ARAPUCA device have also been performed~\cite{Paulucci_2020,Bertolini_XARAPUCAOptimization}. The present study focuses on improving the speed of simulated optical photon propagation through the bulk detector volume. A complete photon simulation of the FD-HD, incorporating X-ARAPUCAs, wavelength shifting, and detailed photon interactions, lies beyond the scope of this work.

Since optical transport is the most computationally demanding stage, we demonstrate photon-by-photon propagation to the X-ARAPUCA surface as a proof of concept. The speedup achieved enables future investigations of wavelength shifting and detector response, where a comprehensive treatment of systematics will be pursued.

%The vertical drift module (FD2-VD) features a different design, with the anodes formed by charge-readout planes (CRPs) at the top and bottom of the active volume, while the cathode is suspended at mid-height, dividing the detector into two vertically stacked drift volumes of 6.5~m each. Each CRP consists of stacked, segmented, and perforated printed circuit boards (PCBs) with etched electrode strips~\cite{DUNE:2023nqi}, the principal departure from the wire-based anodes of FD-HD. This geometry has direct consequences for the PDS: although perforated, the CRP structure is effectively opaque to light, precluding the anode-integrated PDS arrangement used in FD-HD. Photon detectors can therefore be placed only on the cathode plane, or behind the field cage on the cryostat walls, provided the field cage is sufficiently transparent to light. In the FD2-VD design, \textcolor{red}{320} double-sided X-ARAPUCA modules are evenly distributed across the cathode plane, and \textcolor{red}{352} single-sided modules are mounted on the cryostat walls behind the field cage, for a total of 672 modules~\cite{DUNE:2023nqi}.

\section{Photon Simulation}
The simulation of optical photons is a multi-stage process. It begins with the generation of primary particles that traverse the target material, ionizing and exciting the liquid-argon atoms along their paths. The number of scintillation photons produced at each step of the particle trajectory is then computed, and these photons are transported through the detector geometry until they are absorbed or reach a PD. Finally, the PD response is digitized to produce realistic waveforms. Given the size of DUNE's far-detector modules, the photon-transport stage dominates the computational cost, as each photon must be propagated across many meters of active volume; it is this stage that is the focus of the present work. In this paper we consider only photons produced via scintillation, the dominant light-production mechanism in liquid argon. Both ionization and excitation of argon atoms lead, on picosecond timescales, to the formation of Ar$^{*}_{2}$ excimers in singlet or triplet states, whose radiative decays produce a characteristic VUV emission centered at 128~nm~\cite{Doke:2002oab}. The scintillation light yield depends on the applied electric field through electron-ion recombination. Photons are produced via two channels: direct excitation of argon atoms, and recombination of ionization electrons with argon ions, both of which feed the same excimer states. At zero field, where recombination is maximal, liquid argon yields approximately $4.2 \times 10^{4}$ photons per MeV deposited by a minimum-ionizing particle, corresponding to an average energy expenditure of $W_{\mathrm{ph}} \simeq 23.6$~eV per photon~\cite{Doke:2002oab}. An applied drift field sweeps a fraction of the ionization electrons away from their parent ions before they can recombine, suppressing the recombination channel; the light yield is therefore anti-correlated with the collected ionization charge. At the nominal DUNE drift field of 500~V/cm, the recombination luminescence is quenched, and the expected yield is approximately $2.4 \times 10^{4}$ photons per MeV~\cite{DUNE:2020txw}. This anti-correlation also implies that the light and charge signals carry complementary calorimetric information, and their combination can improve the energy resolution~\cite{Doke:2002oab}.

DUNE adopts the Geometry Description Markup Language (GDML)~\cite{GDML} as its canonical geometry source, ensuring interoperability across toolkits: the same GDML file can be consumed directly by both \gfour~and ROOT. Once the \gfour~geometry has been constructed from this description, it is handed to \opticks, which translates the detector geometry into NVIDIA OptiX acceleration structures for GPU-based optical photon propagation. To study the benefits of \opticks~for DUNE geometries, two complementary simulation workflows were developed. The first focuses on stand-alone simulations with \gfour~ and \opticks, allowing a direct comparison of CPU- and GPU-based photon transport. The second integrates \opticks~ within \texttt{LArSoft}, enabling large-scale optical photon studies in the full DUNE software stack. \texttt{LArSoft}, is a set of detector-independent software tools for the simulation, reconstruction, and analysis of data from  LArTPC neutrino experiments, built on the \texttt{art} event-processing framework~\cite{Snider:2017wjd}. % Because the common features of LArTPCs allow algorithm code to be shared across detectors of very different size and configuration, \texttt{LArSoft} is used in production by several experiments, including ArgoNeuT, MicroBooNE, ICARUS, SBND, and DUNE; individual experiments contribute their detector-specific geometry descriptions and electronics models. 
In \larsoft, events are processed through a sequence of modules, each responsible for a specific task in the simulation or reconstruction chain. For example, modules handle primary particle generation, the propagation of particles through detector material using \gfour, and the processing of detector signals, such as the reconstruction of signals from photon detectors. At each \gfour~ step, specialized modules like the \texttt{IonAndScint} are invoked to estimate the amount of scintillation light and ionization charge. This modular design ensures that physics processes are encapsulated within well-defined components, allowing experiments to extend or customize functionality while maintaining compatibility with the broader \larsoft~ framework.
Integrating \opticks~ as a \larsoft~module within this architecture makes GPU‑accelerated optical transport directly available to any experiment using \larsoft, with the present work focused on the implementation for DUNE’s far‑detector horizontal drift configuration. These workflows are discussed in the following subsections.

Both workflows use \texttt{RiceOpticks}, a fork of \opticks~developed for this work~\cite{RiceOpticks}. It introduces several capabilities required for large-scale simulation: accelerated import of large detector geometries; export of global photon-sensor identifiers to the GPU so that hits can be mapped back to individual photon detectors; a custom scintillation-photon collection function that does not rely on \texttt{G4Step} or \texttt{G4Track}; and propagation of parent-particle identifiers alongside each photon. The last of these preserves the truth-level association between detected photons and their originating energy deposits, which makes \texttt{RiceOpticks} compatible with the \texttt{LArSoft} data model.

\subsection{Stand-alone Simulation with \gfour~ and \opticks}\label{subsec:standalone}
\begin{figure}[ht!]
  \includegraphics[width=0.48\textwidth]{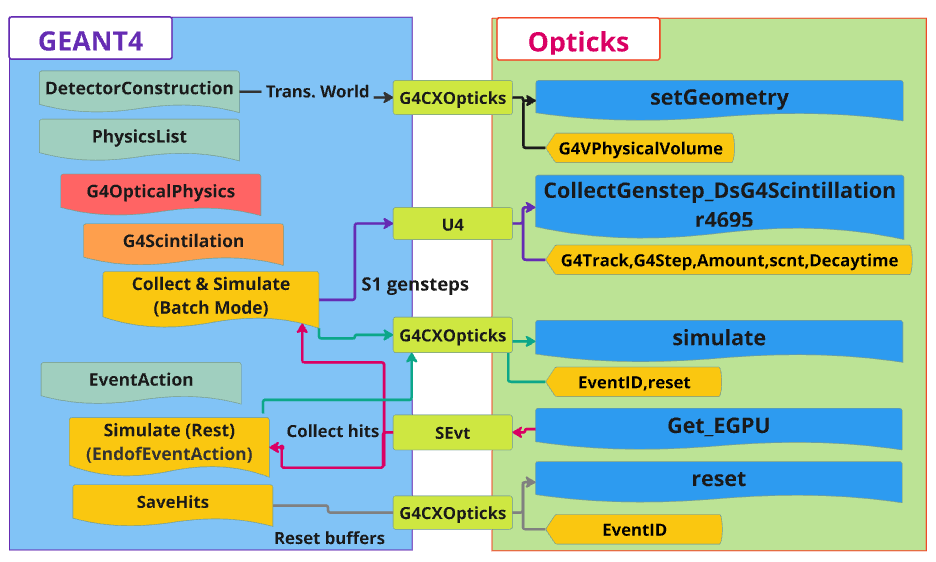}
\caption{The diagram displays how scintillation photon information is propagated from \gfour~ to \opticks~ in the standalone workflow. During event processing, each step that produces scintillation photons invokes photon collection. If the number of collected photons exceeds the maximum photon setting, the photons are simulated in batches; otherwise, they are simulated at the end of the event. Arrows indicate the direction of information flow.}
\label{fig:workflow_sd}  
\end{figure}
For initial testing and validation, a stand‑alone simulation framework was implemented to generate photons from any GDML geometry using both \gfour~ and \opticks. Optical properties were defined in the GDML, with auxiliary tags employed to identify PDs. 
After the GDML loading, a unique sensor identification number (SID) is assigned to each PD. These SIDs are used consistently in both \gfour~ and \opticks~ to simulate scintillation photons and to compare the results between the two frameworks.

\opticks~ relies on \gfour~ to track the steps of charged particles during the scintillation process and to estimate the number of scintillation photons to generate. This metadata is passed directly to \opticks~ before the photon‑generation loops in \texttt{G4Scintillation}, eliminating the need for photon transfer within \texttt{SteppingAction}. Accessing scintillation photons through \texttt{SteppingAction} would require iterating over secondary‑particle buffers after \gfour~ has already created optical secondaries, introducing unnecessary overhead and complexity. By intercepting at the process level instead, \opticks~ operates directly on physical step data, providing a cleaner and more efficient mechanism for GPU‑based photon transport. Photon generation can be configured to occur either in batches by number of photons or as a single comprehensive ensemble, contingent upon the specific requirements of the study. Following each GPU‑based propagation cycle, detector hits are recorded in a ROOT file. The described workflow is illustrated in Fig.~\ref{fig:workflow_sd}.

Within this workflow, a custom primary photon generator was implemented to simulate photons with identical kinematics on both GPU and CPU, enabling direct photon‑by‑photon comparison. Two approaches are supported: the first relies on CPU sampling or input from a ROOT file and transfers photons to the GPU. This method is slower but allows precise photon‑by‑photon validation. The second performs independent sampling directly on the GPU, which is faster since no CPU‑to‑GPU transfer is required. In addition, Gaussian wavelength profiles can be generated, with \gfour~ and \opticks~ handling the sampling independently. This custom primary photon generator was used for both performance evaluation and validation purposes.

\begin{figure}[h!]
  \includegraphics[width=0.48\textwidth]{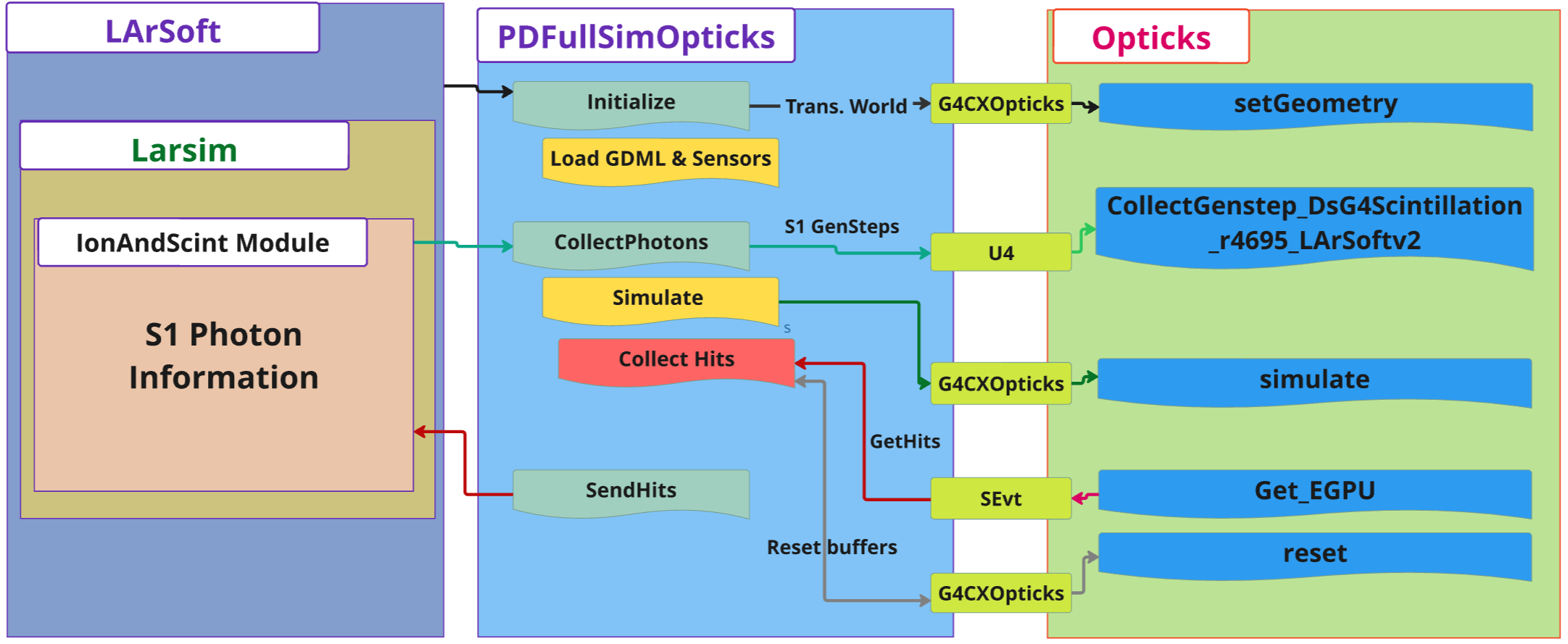}
\caption{ The diagram illustrates the \texttt{PDFullSimOpticks} workflow, which implements optical simulation within \texttt{LArSoft}. The \texttt{IonAndScint} module provides detailed information about scintillating particles, including step data and the number of generated photons. This information is collected by \texttt{PDFullSimOpticks} and transferred to \texttt{RiceOpticks} via a custom function to simulate photons on the GPU. }
\label{fig:workflow_larsoft} 
\end{figure}
\subsection{Opticks implementation in the DUNE software stack}

In DUNE, the simulation workflow begins at the event-generation stage. Reflecting the breadth of the DUNE physics program, this spans energy depositions from a few MeV to tens of GeV, and different event generators (e.g., GENIE~\cite{Andreopoulos:2009rq}, MARLEY~\cite{Gardiner:2021qfr}, and single-particle generation in \gfour) are employed according to the primary interaction: GENIE for beam and atmospheric neutrino interactions and proton decay, MARLEY for low-energy supernova neutrinos, and single-particle generation handled directly by \gfour~for studies of particles traversing the liquid argon. In the second stage, \gfour~transports the final-state particles through the detector, recording energy depositions along their trajectories. For each energy deposition, the number of scintillation photons is computed from the ionization-and-scintillation model described in \cite{Marinho_2022}. Currently, DUNE optical photon simulations rely on alternative methods such as photon library and semi‑analytical approaches rather than full optical simulations due to the large photon processing cost. For our purposes, to maintain full optical simulation, the \texttt{PDFullSimOpticks} module~\cite{laropticks} was developed to interface \opticks~ with \texttt{LArSoft}. This module enables GPU‑accelerated optical photon propagation within the standard \texttt{LArSoft} workflow.

In the current method, as shown in Fig.~\ref{fig:workflow_larsoft}, optical sensors are identified and assigned SIDs, which are provided by \texttt{LArSoft} during the initialization stage. Scintillation-photon information associated with each energy deposit is obtained using the \texttt{IonAndScint} module and subsequently transferred to \opticks~ through the \texttt{PDFullSimOpticks} interface. After the GPU‑based optical simulation is completed, the resulting optical hits are collected for each PD and returned to \texttt{LArSoft} as \texttt{OpDetBacktrackerRecord}, a \texttt{LArSoft} data product that associates each photon hit with its parent particle.

The workflows described above are utilized in the following sections for validation and performance comparison.

\section{Opticks and GEANT4 Comparison}

In this section, we present the physics validation and performance evaluation of the \opticks~integration for DUNE, benchmarked against \gfour.
In this study, absorption lengths and Rayleigh scattering of liquid argon, as well as reflections from metals, were included according to values exist in \texttt{LArSoft}~\cite{OpticalProp}.
The \opticks~simulations were performed on an NVIDIA RTX 5090 GPU, while \gfour~and semi-analytical simulations ran on an AMD Ryzen Threadripper 7970X CPU.

\begin{table}[H]
    \centering
     \begin{tabular}{ p{3.8cm}p{3.0cm}  }
        \hline
        Name & Version \\
        \hline
        \gfour  &  11.1.2 \\
        \opticks & 1.0r \\ 
        \texttt{Cuda} & 13.0 \\
        \texttt{NVIDIA OptiX } & 9.0 \\
        \texttt{NVIDIA Driver} & 580.126.18 \\
        \texttt{DUNESW} & 10\_20\_00d00 \\
        \hline
    \end{tabular}
    \caption{The following packages and their main dependencies were used for simulation validation. }
    \label{tab:packages}
\end{table}

\begin{figure}[]
  \includegraphics[width=0.47\textwidth]{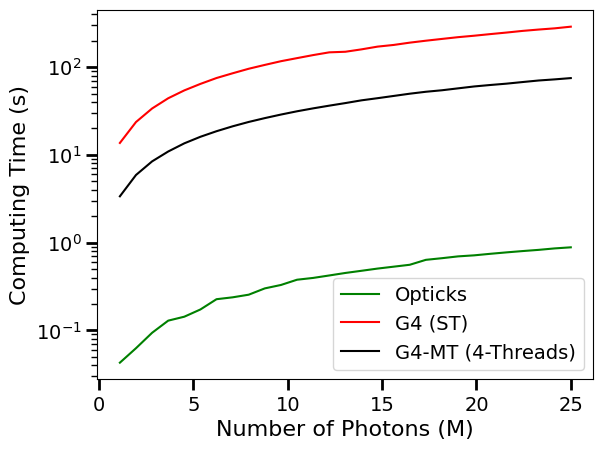}
\caption{Performance curves comparing single-threaded, multi-threaded \gfour, and \opticks.}\label{fig:comparison}       % Give a unique label
\end{figure}

The software stack and their main dependencies used for simulation validation are listed in Table~\ref{tab:packages}.
The details of these comparisons are discussed in the following subsections.

\begin{figure*}[!h]
  \centering
  \includegraphics[width=0.48\textwidth]{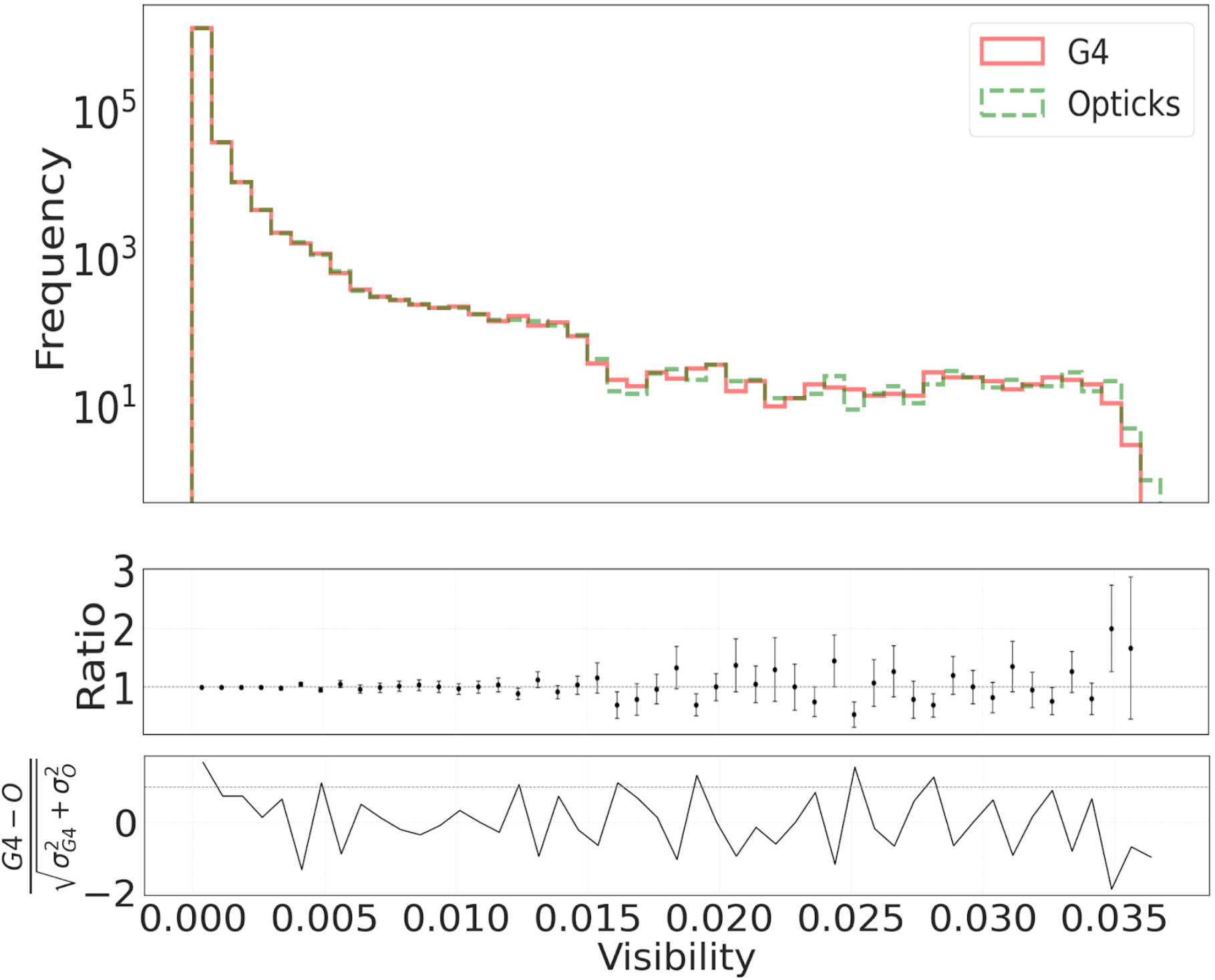}
  \hfill
  \includegraphics[width=0.48\textwidth]{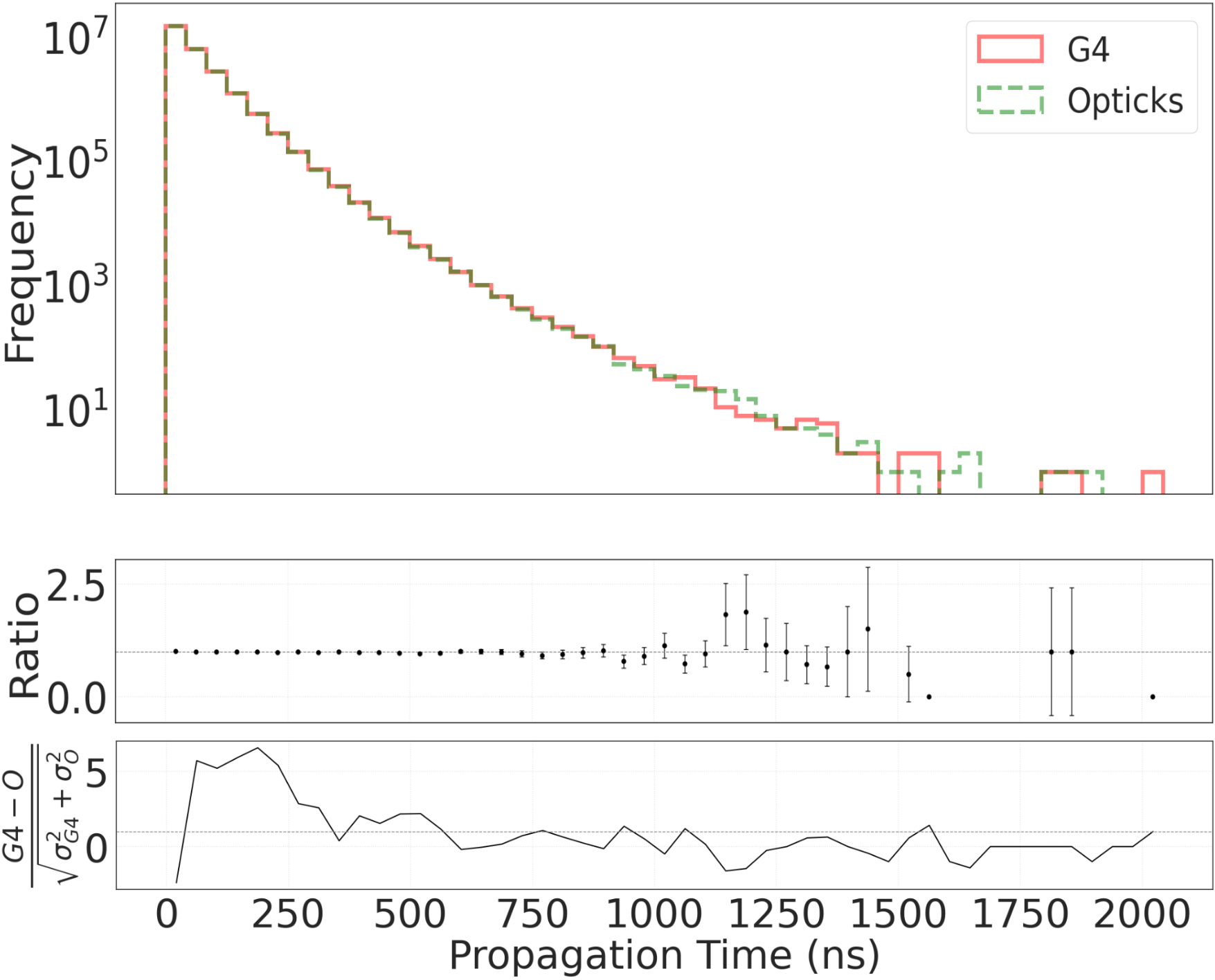}
  \caption{Left: comparison of visibility distributions between \gfour~and \opticks. Right: propagation time of detected photons.}
  \label{fig:optical_comparison}
\end{figure*}
\subsection{Performance Evaluation}

To demonstrate the feasibility and quantify the benefits of integrating \opticks~into the DUNE software stack, 
we conduct a performance benchmark against the reference \gfour~optical simulation using the stand-alone workflow described in Section~\ref{subsec:standalone}. 
For the performance benchmark, the FD-HD geometry with a single liquid-argon volume was used, excluding wires.

Using a custom photon generator, as described in Section~\ref{subsec:standalone}, 
between 0.25 and 25 million photons with a Gaussian energy distribution centered at 9.7~eV 
and a standard deviation of 0.1~eV were simulated isotropically within the detector volume.
The simulations were performed using three configurations: 
\gfour~in single-threaded mode, \gfour~in multi-threaded mode with four threads, 
and \opticks~on a single GPU.
Computation time was measured for all configurations from the start of the event to the end, 
excluding initialization and outputting times.
The results are presented in Fig.~\ref{fig:comparison}, which shows computation time (in seconds) 
as a function of photon count (in millions).
For each performance curve, we determined the slope of the linear fit to the data, 
interpreting it as the processing cost per million photons. Performance improvement was quantified as the ratio of the \gfour~slope to the 
\opticks~slope, yielding speedups of \(313 \pm 3\) in single-thread mode and 
\(83 \pm 1\) with four threads. The uncertainties were obtained from the regression fits of 
the slopes and propagated to the performance ratios.

For GPU runs, resource consumption was monitored across both host system memory and GPU hardware during execution. On the host side, peak system memory usage reached 4.5 GB RAM. On the device side, maximum GPU memory consumption was measured at 5.8 GB VRAM, representing 18\% of the total available video memory. The remaining 82\% (26.2 GB) of unallocated VRAM allows multiple simulation events to be executed concurrently on a single GPU. Under a full processing workload, GPU compute utilization reached a maximum of 100\%, with a corresponding peak power draw recorded at 270 W. 
%After presenting the benchmark results, various validations were performed and are discussed in the following section.

\subsection{Validation}\label{subsec:validation}
To compare optical photon propagation between \opticks{} and \gfour{}, a reduced FD-HD geometry was employed. 
This geometry spans one APA in width, two APAs in height, and six APAs in length, corresponding to an active volume of $719\text{ cm} \times 1208\text{ cm} \times 1394\text{ cm}$. 
The volume was discretized into $17 \times 17 \times 17$ bins, yielding $4{,}913$ voxels with dimensions of approximately $42.3\text{ cm} \times 71.1\text{ cm} \times 82.0\text{ cm}$. 
For each voxel, $10^{5}$ optical photons were generated uniformly throughout the voxel volume using the visibility service in \larsoft{} and stored in a ROOT file. 
Subsequently, the photon information from this ROOT file was imported into the standalone simulation, where photons were propagated with identical kinematics in both \gfour{} and \opticks{}.

Photons reaching the X-ARAPUCA modules, which serve as the sensitive regions, were recorded as optical hits with 100\% efficiency. After photon simulation was completed, visibility fractions as a function of SIDs were estimated according to Eq.~\ref{eq:visibility}.
\begin{equation}
\label{eq:visibility}
    V^{i,j} = \left(\frac{N_{det}^{i}}{N_{sim}}\right)^{j},
\end{equation}
Where $V^{i,j}$ denotes the visibility at sensor $i$ and voxel $j$, $i$ goes from $[0,479]$, a total of 480 sensors. The denominator $N_{sim}$ corresponds to the total number of photons simulated at voxel $j$. Visibilities from all voxels and sensors were compiled into histograms and compared between \gfour~and \opticks, as shown in Fig.~\ref{fig:optical_comparison} left panel. The resulting visibility distributions from \opticks~and \gfour~are in close agreement. Visibilities are used to generate lookup tables, which provide an alternative approach to estimating optical photon hits. Although \opticks~can eliminate the need for lookup tables entirely, it generates them faster, as demonstrated by the propagation time when comparing against \gfour~as shown in Fig.~\ref{fig:comparison}.
\begin{figure}[H]
  \includegraphics[width=0.48\textwidth]{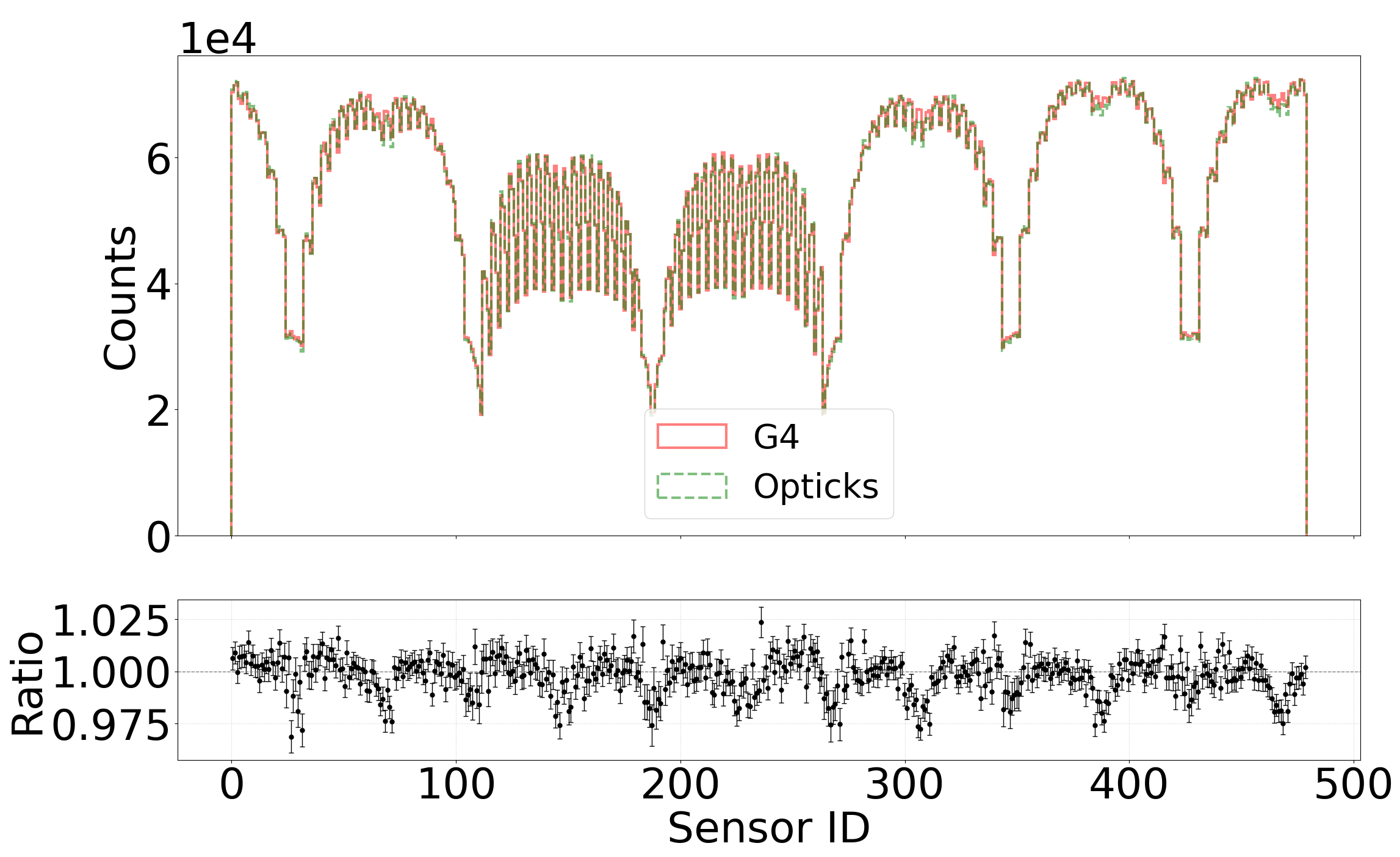}
\caption{Detected photons as a function of sensor identification number (SID).}
\label{fig:sensorID}       % Give a unique label
\end{figure}
Comparing the detected photon counts across SIDs demonstrates agreement between \opticks~and \gfour, as shown in Fig.~\ref{fig:sensorID}. Photon propagation times for detected photons were histogrammed for every voxel and are shown in Fig~\ref{fig:optical_comparison} right panel. This comparison also yields consistent results between \gfour~and \opticks.
%\begin{figure}[h]
%  \includegraphics[width=0.47\textwidth]{figures/propagation_time.png}
%\caption{Propagation Time of Detected Photons}
%\label{fig:proptime}       % Give a unique label
%\end{figure}

%Add Error here
We estimate the uncertainty on the ratio plots $\sigma_{R}$ by following standard error propagation.

\begin{gather}\label{eq:ratio}
    R = \frac{N_{O}}{N_{G4}}, \quad
    \sigma_{N_{G4}} = \sqrt{N_{G4}}, \quad
    \sigma_{N_{O}} = \sqrt{N_{O}}, \\
    \sigma_{R} = R \, \sqrt{\left(\frac{\sigma_{N_{O}}}{N_{O}}\right)^{2}
    + \left(\frac{\sigma_{N_{G4}}}{N_{G4}}\right)^{2}}.
\end{gather}
where $N_{O}$ and $N_{G4}$ denote the number of entries in each histogram bin for the \opticks~ and \gfour~simulations, respectively, with the statistical uncertainties on the bin given by $\sigma_{N_{G4}}$ and $\sigma_{N_{O}}$.

\begin{figure}[H]
  \includegraphics[width=0.47\textwidth]{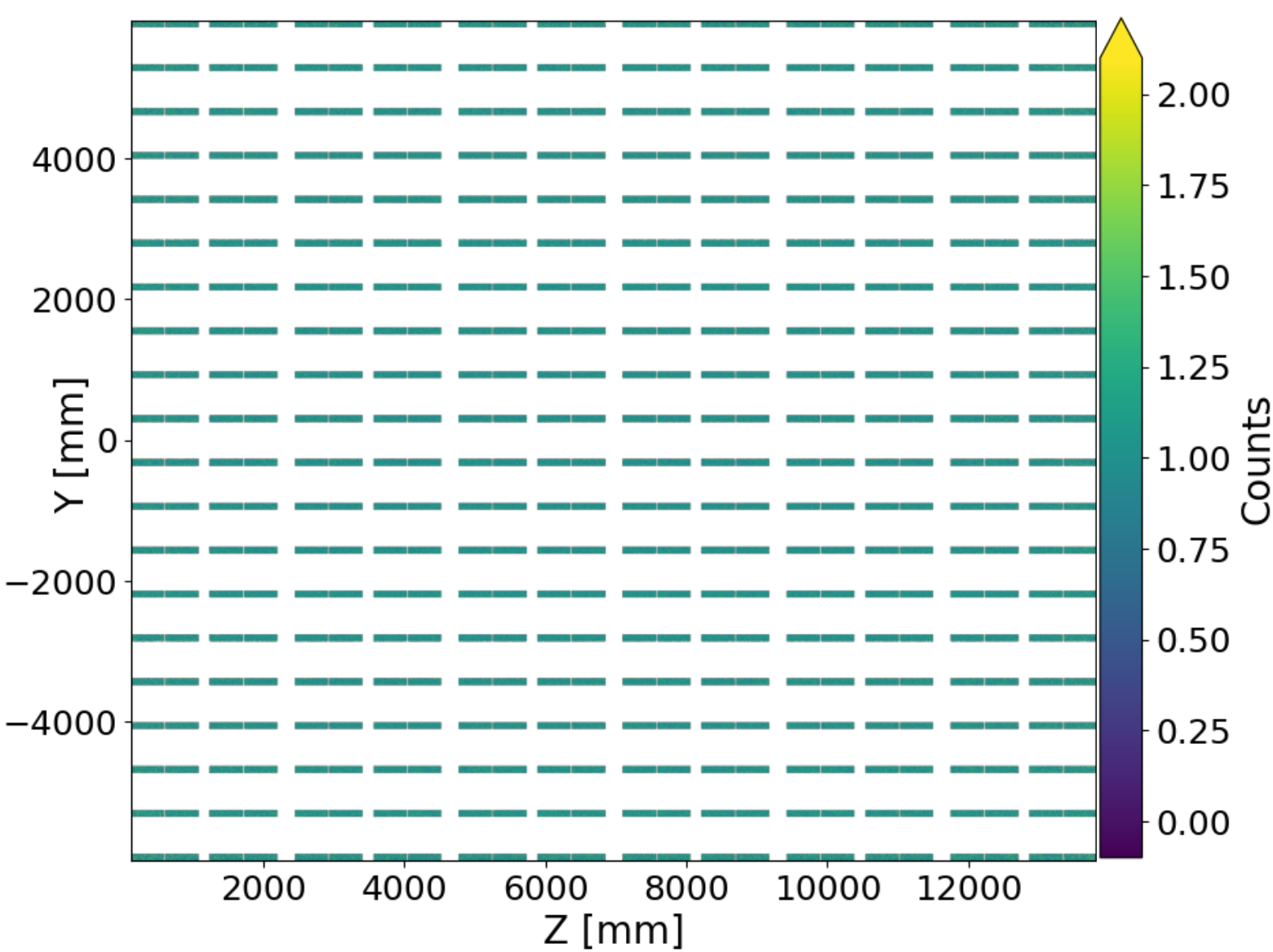}
\caption{Ratio results from 2D Histogram of Y–Z coordinates for optical hits at each photon detector in the reduced geometry. }
\label{fig:hits}       % Give a unique label
\end{figure}

The spatial distribution of optical hits was compared across the $Y$--$Z$ plane by taking the ratio of 2D hit histograms between \opticks~ and \gfour. As shown in Figure~\ref{fig:hits}, the spatial distributions are consistent across the entire active region, confirming agreement in geometry and physics modeling between \opticks~and \gfour.

The agreement observed in 2D spatial hit profiles and time‑of‑flight distributions confirms that \opticks~ reproduces the physics of \gfour~ with high fidelity. This validates the GPU‑accelerated framework as a reliable substitute for full optical simulation across the active region, delivering significant performance improvements. Moreover, this consistency is aligned with results reported in other implementations when comparing with~\gfour~\cite{Li:2023ocy,NEXT:2025opticks,Galgoczi2026}.

%We also evaluated the portability of this across GPU architectures. The benchmark results presented above were obtained on a  GPU card equipped with dedicated ray-tracing (RT) cores, which the OptiX engine exploits for hardware-accelerated geometry intersection. Since such hardware is not always available on production computing resources, we repeated the performance tests on data-center GPUs commonly deployed at HPC and grid sites, including the NVIDIA A100, which lacks dedicated RT cores and performs ray tracing in software, and the more modest NVIDIA L4. While the RT cores provide an additional performance boost, both data-center GPUs nevertheless delivered speedups of \textcolor{red}{[X]}$\times$ and \textcolor{red}{[Y]}$\times$, respectively, over the single-threaded \gfour~baseline. This demonstrates that the gains of GPU-accelerated optical simulation are not contingent on specialized ray-tracing hardware, and that the workflow can be deployed effectively on the heterogeneous GPU resources available to DUNE. Fig~\ref{fig:comparison} shows the comparison of different hardware as explained previously. 

%Motivated by these results, we proceeded to the full integration of \opticks~into the DUNE software stack and a comparison against the currently implemented semi-analytical approach.
\begin{table*}[!t]
\centering
\caption{Comparison of processing time for \opticks~and the
semi-analytical model for
representative DUNE physics interaction samples. The number of photons is proportional to the energy deposited, time is the average time per event. }
\label{tab:interaction_comparison}
\begin{tabular}{lllcc}
\toprule
 & & & PDFullSimOpticks & Semi-analytical \\
\cmidrule(lr){4-4} \cmidrule(lr){5-5}
Interaction type & Average Edep [GeV] & Average \# photons & Time [sec] & Time [sec] \\
\midrule
CC $\nu_e$ interactions                  & 8.43  & $1.13\times10^{7}$ & 3.17  & 3.50 \\
CC $\nu_\mu$ interactions                & 3.76  & $4.79\times10^{6}$ & 1.91 & 1.82  \\
CC $\nu_e$ MARLEY (supernova)            & 0.03 & $3.84\times10^{4}$ & 0.13 & 0.03  \\
Proton decay $p\rightarrow K^{+}\nu$     & 0.33  & $4.55\times10^{5}$ & 0.34 & 0.18  \\
\bottomrule
\end{tabular}
\label{tab:2}
\end{table*}

\section{ PDFullSimOpticks in LArSoft}
%Each optical hit is subsequently digitized to simulate the detector response. Finally, optical \emph{flashes} are reconstructed, where a flash is defined as at least \textcolor{red}{20?} optical hits registered within a \textcolor{red}{100~ns?} time window; flashes provide the event time $t_{0}$, which is essential for non-beam events such as supernova neutrinos and nucleon decay candidates. This workflow is highly flexible: because photon generation is decoupled from particle transport, systematic studies of ionization and light-yield models can be performed without regenerating the final-state particles, and detector-response variations or reconstruction-threshold optimizations can be studied in the same manner. 

After the integration of \opticks~via \texttt{PDFullSimOpticks} into \texttt{LarSoft}, we simulated a representative set of physics interactions: charged-current $\nu_\mu$ and $\nu_e$ interactions from the beam and from atmospheric neutrinos, supernova $\nu_e$ interactions generated with MARLEY, and proton decay events, in the FD-HD reduced geometry. Table~\ref{tab:2} reports the average processing time per event for each interaction type and its corresponding time when using the semi-analytical approach currently implemented as the default method in the DUNE software stack.
Processing times are obtained from \texttt{TimeTracker} module within \texttt{LArSoft}.
These results demonstrate that the full optical simulation with \opticks~achieves processing times comparable to the semi-analytical approach. As the number of photons increases, \opticks~simulates them more efficiently than the semi-analytical method. 
%A full optical simulation at this cost enables systematic studies that were previously prohibitive, while retaining the complete truth-level backtracking of each detected photon to its originating energy-deposit information.

\section{Conclusion}

In this work, we have presented the first performance evaluation and implementation of GPU-accelerated optical photon simulation in a 10~kt-scale detector, applied to the DUNE far-detector horizontal drift configuration. A stand-alone benchmark on the full FD-HD geometry shows that \opticks\ simulates the same number of photons in comparison to a single-thread \gfour, with a speedup of 313 $\pm$ 3, and 83 $\pm$ 1 when comparing to a four-thread. Photon transport in \opticks~ was validated against the reference \gfour~ simulation using a reduced FD-HD geometry, demonstrating agreement across the metrics detailed in Sec.~\ref{subsec:validation}.
Building on these results, we developed \texttt{PDFullSimOpticks}, a dedicated module that integrates \opticks~into the \texttt{LArSoft}-based DUNE software stack. Making the full chain, from event generation through photon transport, optical-hit digitization, and flash reconstruction, possible within DUNE software stack. Thanks to the integration being implemented at the \texttt{LArSoft} level rather than being specific to DUNE, GPU-accelerated optical transport becomes directly available to any experiment that uses \texttt{LArSoft}.

Crucially, and unlike parameterized or surrogate methods, the \opticks~simulation captures true stochastic fluctuations by propagating every photon individually with \gfour-equivalent physics, preserving the complete truth-level backtracking of each detected photon to its originating energy deposit and parent particle. This combination of full fidelity and tractable cost establishes a new paradigm for the optical simulation of kiloton-scale detectors, studies that were historically precluded by the computational burden of photon transport, high-statistics optimization of photon-detector layouts, data-driven tuning of optical models, systematic uncertainty quantification for light-based calorimetry, and trigger studies for low-energy signals such as supernova bursts will be feasible. Moreover, the ability to generate large volumes of high-fidelity, fully labeled optical data at GPU speed directly addresses one of the principal bottlenecks for machine learning in this domain: conventional CPU-based simulation could not produce training datasets of the required scale, while fast methods lack the truth information needed for supervised learning and validation. The AI-ready datasets enabled by this work open the door to simulation-based inference, ML-driven event reconstruction, flash matching, detector calibration, and the training of surrogate models that are themselves anchored to a full simulation. 

\section{Acknowledgments}
This work was supported by U.S Department of Energy, Office of Science, Office of High Energy Physics, DE-SC0026350. The authors extend their sincere appreciation to Seth Johnson, Soon Yung Jun,  Erica Snider, and Stefano Tognini, for useful discussions regarding the integration of \opticks~into \texttt{LArSoft}.

\bibliographystyle{spphys}       % APS-like style for 
\bibliography{references}   % name your BibTeX data base

\end{document}